\documentclass[11pt]{article}
\usepackage[margin=1in]{geometry}
\usepackage[T1]{fontenc}
\usepackage{lmodern}
\usepackage{amsmath,amssymb}
\usepackage{booktabs}
\usepackage{graphicx}
\usepackage{tikz}
\usetikzlibrary{arrows.meta, positioning, shapes.geometric}
\usepackage{pgfplots}
\pgfplotsset{compat=1.16}
\tikzset{
  vbox/.style={rectangle, rounded corners=2pt, draw=black!70, fill=black!4,
    minimum height=7.5mm, minimum width=28mm, align=center, font=\small},
  vspec/.style={vbox, fill=blue!8, draw=blue!50!black!70},
  vdec/.style={diamond, aspect=2.4, draw=black!70, fill=black!4, align=center,
    font=\small, inner sep=1pt},
  varr/.style={-{Stealth[length=2.2mm]}, semithick, black!75},
  vlab/.style={font=\scriptsize, black!70}
}
\usepackage{multirow}
\usepackage[hidelinks]{hyperref}
\usepackage{xcolor}
\usepackage{enumitem}
\usepackage{microtype}

\newcommand{\pp}{\,\mathrm{pp}}

\newcommand{\leni}{\textsc{Leni}}

\title{\textbf{Where Does Agent Reliability Come From?\\
\large A Cross-Benchmark Decomposition of Verification Loops, Specialist Models,\\ and Scaffolding in a Production Enterprise Agent}}

\author{Arunabh Dastidar and the Leni Team\\
Leni Inc.\\
Correspondence: \texttt{arunabh@leni.co}}

\date{July 2026 (preprint) \quad evaluations conducted March--April 2026}

\begin{document}
\maketitle

\begin{abstract}
Multi-step enterprise agent tasks fail in a characteristic way: single-pass inference has no checkpoint between deciding an answer and committing to it. We study one production system (\leni{}, an AI business analyst) whose architecture installs such checkpoints: \emph{verification loops} (execute, observe, compare, correct) staffed by lightweight task-specialized post-trained models. We evaluate the unmodified production configuration on three public benchmarks stressing distinct failure modes: SpreadsheetBench Verified (silent computation error), BullshitBench~v2 (premise confabulation), and the GAIA validation split (cascade error over long tool chains). The full system improves over its frontier base model by $+11.0\pp$ on SpreadsheetBench ($91.25\%$ vs.\ $80.25\%$, $n{=}400$, $p<0.001$), $+7$ to $+10\pp$ on BullshitBench ($98\%$ vs.\ $91\%$, $n{=}100$, nominally significant), and ${\sim}{+}15\pp$ on GAIA validation ($75.2\%$ pass@1 vs.\ ${\sim}60\%$, $n{=}165$; $83.0\%$ best-of-$k$). The GAIA figures correct an earlier company report whose $77.6\%$ mixed single-attempt and best-of-$k$ selection across tiers; re-grading every stored trajectory with the official scorer (validated at 100\% agreement against the 218 previously labeled runs) yields the numbers here. Our central contribution, however, is a \emph{decomposition} of that uplift: most of it comes from scaffolding, routing, and specialist models rather than from the verification step itself, whose isolated contribution is small ($+1.5\pp$ measured on SpreadsheetBench) but concentrated exactly at the top of the score distribution, where it converts otherwise-failing tasks. We instrument the deterministic loop end-to-end, yielding an empirical verifier confusion matrix (task-level catch rate $c\approx0.20$, fix rate $r\approx0.75$, no observed false-alarm regressions) that grounds a compounding-reliability model extended to include verifier false alarms. Preliminary specialist-swap ablations suggest the loop's value depends on \emph{who observes}: replacing the small trained verifier with the generating frontier model eliminates most rescues, consistent with self-assessment biases reported in prior work. A valid-premise control (100 legitimate expert-level questions through the same firewall-active harness) shows zero over-rejections, bounding the firewall's false-positive rate at ${\lesssim}3.6\%$ on that distribution. We report all results with confidence intervals, state which cross-system comparisons are and are not statistically resolvable, and enumerate the experiments (an adversarially matched valid-premise control, hidden-test-set submission, external scaffold baselines) required before stronger claims can be made. This is a vendor evaluation of the vendor's own system; the limitations section says so and lists what independent replication would require.
\end{abstract}


\section{Introduction}

Enterprise deployment of language-model agents fails in a characteristic way. A single tool call works. A two-step chain usually works. By the fifth step, or the first fabricated premise, or the first silently miscomputed cell, the system produces a fluent, confident, and wrong result. In consumer chat this is tolerable; in financial close, contract review, diligence, or reconciliation, a confidently wrong output propagates into real decisions and real liability. The difference is one of kind rather than degree: consumer applications can absorb a per-interaction error rate that back-office work cannot, because enterprise outputs feed downstream filings, models, and contracts where errors compound instead of dissipating. Closing that gap is more than a per-application prompt-engineering exercise. It calls for a reliability layer with the character of infrastructure: built once, shared across task families, and depended on the way applications depend on a database's transactional guarantees.

A common response is to wait for a stronger base model. This paper examines a different lever: the architecture wrapped around the model. We study one production system whose design centers on \emph{verification loops}, a control structure that executes an action, independently observes the result, compares the observation against intent, and corrects before committing. Each loop stage is assigned to the lightest model that can perform it reliably, including small ($0.5$--$3$B) post-trained specialists.

Prior work has shown both that structured feedback can improve LLM outputs \cite{madaan2023selfrefine,shinn2023reflexion,gou2024critic,wang2023selfconsistency} and, pointedly, that models are poor at correcting their own reasoning without external signals \cite{huang2024cannot}. We do not try to settle that debate in general. We measure, inside a single production system under identical conditions, \emph{where} the reliability of a deployed agent comes from, and find three things:

\begin{enumerate}[leftmargin=1.5em]
  \item \textbf{Total architectural uplift is large and, on two of three benchmarks, statistically unambiguous} relative to the bare base model ($+11.0\pp$ on SpreadsheetBench, $n{=}400$; $+7$ to $+10\pp$ on BullshitBench, $n{=}100$; ${\sim}{+}15\pp$ on GAIA validation, $n{=}165$). The comparison class matters: an agentic system (tools, multiple inference passes, multiple models) against raw single-pass inference. This measures the value of the architecture as a whole rather than of any single mechanism (\S\ref{sec:results}).
  \item \textbf{Decomposition shows the verification step's isolated contribution is small but positionally decisive.} On SpreadsheetBench, scaffolding and prompting account for $+9.5\pp$ of the $+11.0\pp$ uplift; the deterministic recalculation loop adds the final $+1.5\pp$ by rescuing 6 tasks, which is the difference between a mid-leaderboard and a near-top result. On GAIA, after correcting the headline score to a pre-specified pass@1 (\S\ref{sec:gaia-results}), the loop's marginal contribution over the estimated structure tiers is ${\sim}{+}1\pp$ and cannot yet be cleanly isolated (\S\ref{sec:decomposition}).
  \item \textbf{Who observes matters as much as whether observation happens.} In preliminary specialist-swap runs, moving the observe/compare stage from a small post-trained verifier back onto the frontier model that generated the artifact reduces SpreadsheetBench rescues from 6 tasks to 2, and reduces BullshitBench correct rejection by 4--5$\pp$. This is consistent with self-preference and self-assessment biases documented for LLM evaluators \cite{panickssery2024selfrecognition,zheng2023judging}, and suggests that what carries the effect is the independence of the observer, not just the presence of a loop (\S\ref{sec:decomposition}).
\end{enumerate}

We also contribute (i) a formalization of the verification loop with a taxonomy of verification \emph{oracles} (deterministic, self-reflective, planner-mediated) and a compounding-reliability model extended to account for verifier false alarms (\S\ref{sec:framework}); (ii) the first end-to-end instrumentation of a production deterministic loop, yielding an empirical verifier confusion matrix (\S\ref{sec:sb-results}); and (iii) an explicit accounting of which of our cross-system comparisons are statistically resolvable at current sample sizes, and which are not (\S\ref{sec:gaia-results}, \S\ref{sec:limitations}).

\paragraph{What this paper does not claim.} We do not claim that verification loops alone explain the measured uplift; our own decomposition shows otherwise. We do not claim statistically significant superiority over other agentic systems on GAIA; the differences are within confidence intervals against reconstructed competitor figures. On the self-reflective firewall, we report a measured valid-premise control: 100 legitimate, expert-level questions run through the same firewall-active production harness produced zero refusal-like outputs (\S\ref{sec:bb-results}); what remains open is a control styled adversarially after the benchmark's own deception techniques. These boundaries are drawn deliberately: the durable finding is the decomposition, not a leaderboard position.

\section{Related Work}
\label{sec:related}

\paragraph{Self-correction and its limits.} Self-Refine \cite{madaan2023selfrefine} and Reflexion \cite{shinn2023reflexion} show that iterated self-feedback can improve outputs on some tasks; self-consistency \cite{wang2023selfconsistency} improves reliability by sampling and voting. Huang et al.\ \cite{huang2024cannot} demonstrate that LLMs largely cannot correct their own \emph{reasoning} without external information, and that naive self-critique can degrade performance. Our results are consistent with both lines: our deterministic loop supplies exactly the external signal Huang et al.\ identify as necessary (a recomputation engine), and our specialist-swap ablation suggests that even with external signals, a generator asked to evaluate its own artifact underperforms an independent observer, echoing self-preference effects in LLM evaluators \cite{panickssery2024selfrecognition}.

\paragraph{Tool-grounded verification.} CRITIC \cite{gou2024critic} interleaves generation with tool-based critique; ReAct \cite{yao2023react} interleaves reasoning and acting; AlphaCodium \cite{ridnik2024alphacodium} shows test-based iteration dominating single-shot generation for code. We extend this direction in three ways: we isolate the \emph{observation} step (independent read-back through a separate code path) as the load-bearing component; we organize instantiations by their verification oracle; and we measure the pattern in one production system across three unrelated benchmarks rather than per-task-tuned setups.

\paragraph{Neurosymbolic verification.} Deterministic engines that re-execute artifacts represent the strongest oracle class. On SpreadsheetBench the current leader is a neurosymbolic system with a custom spreadsheet runtime. We show a commodity deterministic oracle (LibreOffice headless recalculation) recovers most of that benefit at far lower engineering cost, and we quantify the residual gap attributable to LLM-mediated (rather than symbolically enforced) comparison (\S\ref{sec:discussion}).

\paragraph{LLM-as-judge.} BullshitBench is scored by an LLM judge panel. Judge bias, including intra-family preference, is documented \cite{zheng2023judging,panickssery2024selfrecognition}; our panel spans three providers, but two of our configurations are Claude-based and one judge is a Claude model. We treat judge-based scores as noisier than exact-match scores throughout and flag this in \S\ref{sec:limitations}.

\paragraph{Benchmarks.} SpreadsheetBench \cite{ma2024spreadsheetbench} measures exact-match spreadsheet manipulation; we use the expert-annotated Verified subset (400 tasks). GAIA \cite{mialon2023gaia} measures long-horizon tool orchestration with exact-match scoring across three difficulty tiers. BullshitBench~v2 \cite{gostev2026bullshit} measures whether a system rejects fabricated premises rather than elaborating on them; we note it is a community benchmark ($n{=}100$) and weight conclusions accordingly. The three were chosen because they stress unrelated failure modes (silent computation error, sycophantic confabulation, cascade error), so an architectural pattern that helps all three is evidence of generality.

\section{The Verification Loop: Definition and a Reliability Model}
\label{sec:framework}

\subsection{Definition}

A \emph{verification loop} is a control structure wrapped around a base-model action with four stages:

\begin{enumerate}[leftmargin=1.5em]
  \item \textbf{Execute}: the base model performs the action (edit a workbook, answer a question, run a tool step).
  \item \textbf{Observe}: the system produces an independent observation of the result, through a different code path than the one that generated it.
  \item \textbf{Compare}: the observation is checked against the original intent (task instruction, plan, or premise set).
  \item \textbf{Correct}: on a detected discrepancy, the system issues a targeted fix and re-observes; on consistency, it commits.
\end{enumerate}

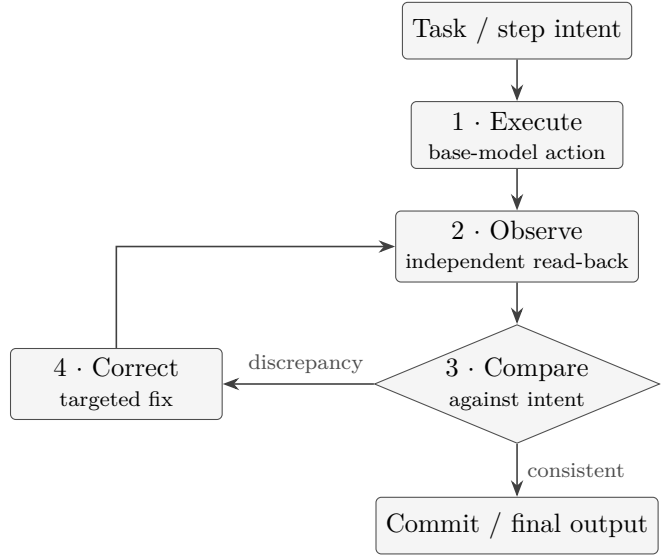
\begin{figure}[t]
\centering
\begin{tikzpicture}[node distance=5.5mm and 14mm]
\node[vbox] (intent) {Task / step intent};
\node[vbox, below=of intent] (exec) {1 $\cdot$ Execute\\[-1pt] {\scriptsize base-model action}};
\node[vbox, below=of exec] (obs) {2 $\cdot$ Observe\\[-1pt] {\scriptsize independent read-back}};
\node[vdec, below=of obs] (cmp) {3 $\cdot$ Compare\\[-1pt] {\scriptsize against intent}};
\node[vbox, left=20mm of cmp] (fix) {4 $\cdot$ Correct\\[-1pt] {\scriptsize targeted fix}};
\node[vbox, below=7mm of cmp] (commit) {Commit / final output};
\draw[varr] (intent) -- (exec);
\draw[varr] (exec) -- (obs);
\draw[varr] (obs) -- (cmp);
\draw[varr] (cmp) -- node[vlab, right]{consistent} (commit);
\draw[varr] (cmp) -- node[vlab, above, pos=0.45]{discrepancy} (fix);
\draw[varr] (fix) |- (obs);
\end{tikzpicture}
\caption{The general verification loop. Observation (stage 2) is what makes defects that are invisible to the generating process visible; correction (stage 4) re-enters observation, not execution.}
\label{fig:loop}
\end{figure}

The load-bearing stage is \emph{observation}. Many defects (mutation-order artifacts, serialization corruption, a formula that evaluates differently than written, a premise the model half-noticed was false) are structurally invisible to the process that created them, and become observable only when the output is re-read through an independent path. The rest of the loop exists to force and act on that observation. Figure~\ref{fig:loop} shows the structure; correction re-enters observation rather than the original execution path.

\subsection{A compounding-reliability model with imperfect verification}
\label{sec:model}

Consider a task of $n$ dependent steps, each succeeding independently with probability $p$. Without verification, end-to-end reliability is $R_{\text{base}} = p^n$, which decays geometrically: at $p=0.95$, a ten-step task succeeds only ${\sim}60\%$ of the time.

A per-step verification loop is characterized by four empirical parameters: the \emph{catch rate} $c$ (probability a true error is flagged), the \emph{fix rate} $r$ (probability a flagged true error is repaired), the \emph{false-alarm rate} $f$ (probability a correct step is flagged), and the \emph{breakage rate} $b$ (probability a false-alarm ``correction'' damages a correct step). Effective per-step reliability becomes
\begin{equation}
p' \;=\; \underbrace{p\,(1 - f b)}_{\text{correct steps that survive}} \;+\; \underbrace{(1-p)\,c\,r}_{\text{errors caught and fixed}},
\label{eq:pprime}
\end{equation}
and end-to-end reliability $R_{\text{loop}} = (p')^n$. The loop helps if and only if $(1-p)\,c\,r > p\,f\,b$: an imperfect verifier can \emph{reduce} reliability if it frequently second-guesses correct work and its corrections are destructive. Because a beneficial correction acts before the error propagates, the benefit compounds multiplicatively, so the loop's marginal value grows with chain length $n$.

Two consequences follow. The model predicts that loops matter most on long dependency chains; our data does not yet test this prediction cleanly (\S\ref{sec:gaia-results}, \S\ref{sec:decomposition}), and we state it as a falsifiable hypothesis rather than a finding. It also makes the verifier's error profile a first-class quantity, and \S\ref{sec:sb-results} reports what appears to be the first published empirical estimate of $(c, r, f)$ from a production deterministic loop.

\subsection{The oracle determines the ceiling}

The compare stage requires a source of truth: the \emph{verification oracle}. It is the most consequential design choice because it bounds $c$ (what can be caught) and $f$ (what is spuriously flagged). We distinguish three classes (Table~\ref{tab:oracles}).

\begin{table}[h]
\centering\small
\begin{tabular}{@{}p{2.3cm}p{3.6cm}p{2.4cm}p{2.6cm}p{3.2cm}@{}}
\toprule
\textbf{Oracle class} & \textbf{Source of truth} & \textbf{Ceiling} & \textbf{Cost} & \textbf{Instantiation} \\
\midrule
Deterministic & External engine re-executes the artifact & Highest (near-symbolic) & Medium (commodity tooling) & Spreadsheet recalculation loop (\S\ref{sec:inst-det}) \\
Self-reflective & Model's structured re-evaluation & Moderate (no external oracle) & Low (protocol only) & Epistemic firewall (\S\ref{sec:inst-self}) \\
Planner-mediated & Typed intermediate artifacts checked against a plan & High on long chains & Medium (planner--executor split) & GAIA re-planning loop (\S\ref{sec:inst-plan}) \\
\bottomrule
\end{tabular}
\caption{Three verification-oracle classes. The oracle sets the loop's reliability ceiling; the loop structure is shared.}
\label{tab:oracles}
\end{table}

\subsection{The model mix: specialists inside the loop}
\label{sec:mix}

Calling a frontier model at every loop stage is expensive and, for observe/compare, potentially counterproductive: the model that produced an artifact is primed to rationalize it \cite{panickssery2024selfrecognition,huang2024cannot}. \leni{} therefore assigns each stage to the lightest model that performs it reliably, post-training small specialists for stages general models do poorly (Table~\ref{tab:specialists}). Specialists are $0.5$--$4$B parameters, cheap enough to run on every iteration. All four are post-trained from open-weight Qwen3 bases (0.6B / 1.7B / 4B, Apache-2.0) with a shared recipe: distillation SFT from a Claude Opus 4.6 teacher (labels grounded in the strongest available oracle, rationales from the teacher), followed by task-specific reinforcement (RLVR/GRPO where a deterministic checker exists, for Cell-S, Parse-S, and Route-XS; SFT-then-DPO on preference pairs for Triage-S), 4-bit quantization for serving, and shadow deployment behind the live loop before promotion. Training data is Leni's production corpus: three years of enterprise back-office tasks (financial institutions and real estate operators) and user accept/correct/reject judgments, collected independently of any benchmark campaign, plus synthetic augmentation derived from production scenario patterns. The authors attest that no benchmark items, and no corpora styled after the benchmarks, were used in training; the verification status of this attestation is discussed in \S\ref{sec:contamination}. Weights and training data are proprietary; \S\ref{sec:limitations} lists this among the obstacles to independent replication.

\begin{table}[h]
\centering\small
\begin{tabular}{@{}llp{5.2cm}p{3.1cm}l@{}}
\toprule
\textbf{Specialist} & \textbf{Size} & \textbf{Post-training objective} & \textbf{Loop role} & \textbf{Used in} \\
\midrule
Leni-Cell-S & ${\sim}4$B & Spreadsheet cell-diff verification (recomputed value vs.\ intent) & Observe / Compare & Deterministic loop \\
Leni-Triage-S & ${\sim}4$B & Premise decomposition + epistemic validity classification & Observe / Compare & Self-reflective loop \\
Leni-Parse-S & ${\sim}1.5$B & Typed-artifact extraction from raw tool output & Observe (typing) & Planner-mediated loop \\
Leni-Route-XS & ${\sim}0.5$B & Step-type classification for per-step routing & Dispatch / route & Planner-mediated loop \\
\bottomrule
\end{tabular}
\caption{The lightweight post-trained specialist mix. Because specialists run on every iteration, their cost and latency matter as much as their accuracy (\S\ref{sec:cost}).}
\label{tab:specialists}
\end{table}

\section{Three Instantiations}
\label{sec:instantiations}

All three instantiations run inside the same production system with no benchmark-specific modifications; they differ only in the oracle. One caveat before the details: while the \emph{harness} is unmodified, individual components were built for production task families that overlap with what these benchmarks measure (a spreadsheet verifier is, unavoidably, aligned with a spreadsheet benchmark). The defensible claim is therefore not ``untuned'' but ``not tuned \emph{to these benchmarks}'': the configurations are the ones serving production traffic, frozen before evaluation (\S\ref{sec:contamination}).

\subsection{Deterministic oracle: the recalculation loop}
\label{sec:inst-det}

Spreadsheet editing is unforgiving: one wrong cell fails the task, and formula engines (Excel, LibreOffice, \texttt{openpyxl}) can evaluate the same formula differently, so a value that looks correct at write time can be silently wrong. The generating model is blind to this because \texttt{openpyxl} does not compute formulas.

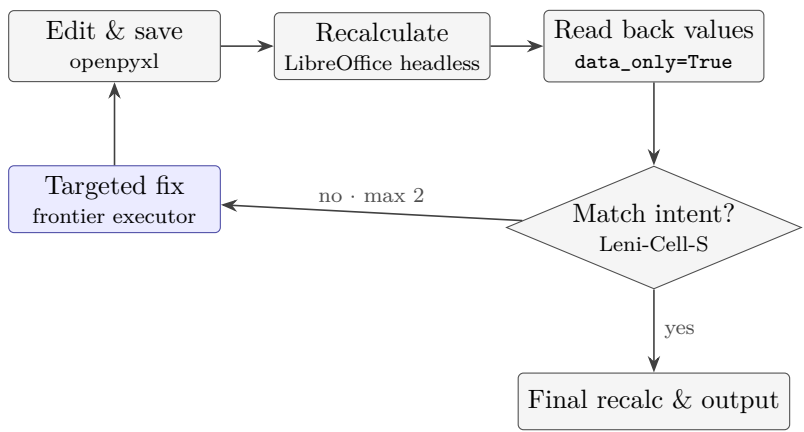
\begin{figure}[t]
\centering
\begin{tikzpicture}[node distance=5mm and 7mm]
\node[vbox] (edit) {Edit \& save\\[-1pt] {\scriptsize openpyxl}};
\node[vbox, right=of edit] (recalc) {Recalculate\\[-1pt] {\scriptsize LibreOffice headless}};
\node[vbox, right=of recalc] (read) {Read back values\\[-1pt] {\scriptsize \texttt{data\_only=True}}};
\node[vdec, below=11mm of read] (match) {Match intent?\\[-1pt] {\scriptsize Leni-Cell-S}};
\node[vspec, below=11mm of edit] (fixit) {Targeted fix\\[-1pt] {\scriptsize frontier executor}};
\node[vbox, below=11mm of match] (out) {Final recalc \& output};
\draw[varr] (edit) -- (recalc);
\draw[varr] (recalc) -- (read);
\draw[varr] (read) -- (match);
\draw[varr] (match) -- node[vlab, right]{yes} (out);
\draw[varr] (match) -- node[vlab, above]{no $\cdot$ max 2} (fixit);
\draw[varr] (fixit) -- (edit);
\end{tikzpicture}
\caption{The deterministic recalculation loop. LibreOffice supplies external ground truth about formula evaluation; the small trained verifier (Leni-Cell-S, shaded) performs the comparison instead of the model that wrote the workbook.}
\label{fig:recalc}
\end{figure}

The loop closes this gap: (a) a frontier executor (Claude Opus 4.6) edits and saves the workbook; (b) the server recalculates all formulas with LibreOffice headless; (c) values are read back through a separate deserialization path (\texttt{openpyxl}, \texttt{data\_only=True}); (d) Leni-Cell-S compares recalculated non-empty cells against task intent and flags discrepancies; (e) on a flag, the frontier executor issues a targeted fix and Leni-Cell-S re-verifies. The loop runs at most two iterations, followed by a final recalculation pass so every formula cell carries a cached value for evaluation (Figure~\ref{fig:recalc}).

\subsection{Self-reflective oracle: the epistemic firewall}
\label{sec:inst-self}

Some tasks have no external oracle: no API returns ``this methodology is fabricated.'' Here the loop uses structured re-evaluation as the oracle. Before answering, the agent must: (1) decompose the question into constituent claims; (2) classify each claim as known-valid, plausible-but-unrecognizable, or real-but-misapplied (both stages on Leni-Triage-S, trained to emit a hard accept/reject per claim rather than continue a fluent answer); (3) on a fabricated claim, produce a \emph{constructive rejection} on the frontier model: name the fabrication, explain why, and redirect to the legitimate concept the user likely intended.

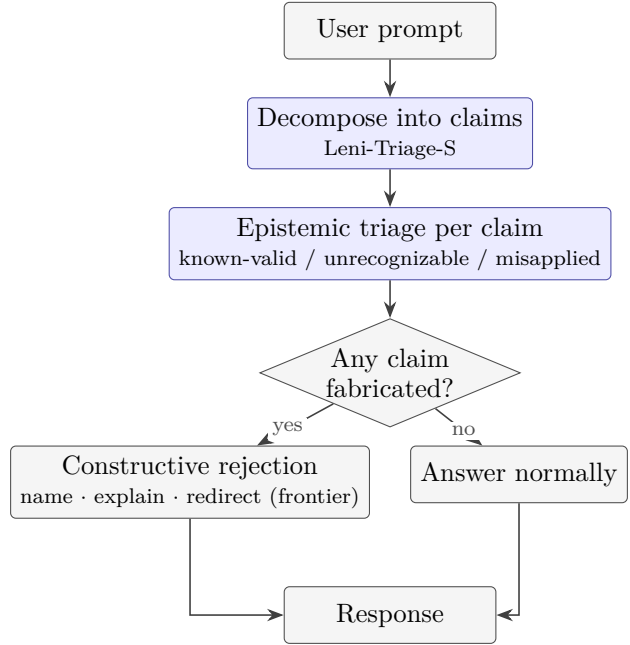
\begin{figure}[t]
\centering
\begin{tikzpicture}[node distance=5mm and 9mm]
\node[vbox] (prompt) {User prompt};
\node[vspec, below=of prompt] (dec) {Decompose into claims\\[-1pt] {\scriptsize Leni-Triage-S}};
\node[vspec, below=of dec] (triage) {Epistemic triage per claim\\[-1pt] {\scriptsize known-valid / unrecognizable / misapplied}};
\node[vdec, below=of triage] (fab) {Any claim\\[-1pt] fabricated?};
\node[vbox, below left=6mm and -6mm of fab] (rej) {Constructive rejection\\[-1pt] {\scriptsize name $\cdot$ explain $\cdot$ redirect (frontier)}};
\node[vbox, below right=6mm and -6mm of fab] (ans) {Answer normally};
\node[vbox, below=21mm of fab] (resp) {Response};
\draw[varr] (prompt) -- (dec);
\draw[varr] (dec) -- (triage);
\draw[varr] (triage) -- (fab);
\draw[varr] (fab) -- node[vlab, fill=white, inner sep=1.5pt, pos=0.6]{yes} (rej);
\draw[varr] (fab) -- node[vlab, fill=white, inner sep=1.5pt, pos=0.6]{no} (ans);
\draw[varr] (rej) |- (resp);
\draw[varr] (ans) |- (resp);
\end{tikzpicture}
\caption{The self-reflective epistemic firewall. With no deterministic oracle available, structured re-evaluation of premises serves as the comparison stage; the shaded stages run on the small trained specialist.}
\label{fig:firewall}
\end{figure}

The mechanism (Figure~\ref{fig:firewall}) targets a specific observed failure: in raw single-pass failures the model often already encodes weak recognition of the fabrication (buried hedges, mid-answer caveats) but has no mechanism to promote that signal from a footnote to a behavioral rejection. The firewall supplies that mechanism.

 Its principal risk is symmetric: a system trained to reject premises may over-reject legitimate but unusual ones. Measuring that false-positive rate requires a control set of valid-premise questions, which we have not yet run; until then, the firewall's \emph{calibration} is unestablished even where its \emph{sensitivity} is high (\S\ref{sec:limitations}).

\subsection{Planner-mediated oracle: re-planning over typed artifacts}
\label{sec:inst-plan}

\begin{figure}[t]
\centering
\begin{tikzpicture}[node distance=5mm and 10mm]
\node[vbox] (task) {Task};
\node[vbox, below=of task] (plan) {Planner\\[-1pt] {\scriptsize owns trajectory}};
\node[vspec, right=of plan] (route) {Dispatch typed step\\[-1pt] {\scriptsize routed by Leni-Route-XS}};
\node[vbox, below=of route] (execu) {Executor\\[-1pt] {\scriptsize routed model + tools}};
\node[vspec, below=of execu] (art) {Typed artifact\\[-1pt] {\scriptsize value $\cdot$ citation $\cdot$ error class (Leni-Parse-S)}};
\node[vdec, left=of art] (chk) {Consistent\\[-1pt] with intent?};
\node[vbox, below=7mm of chk] (final) {Final answer};
\draw[varr] (task) -- (plan);
\draw[varr] (plan) -- (route);
\draw[varr] (route) -- (execu);
\draw[varr] (execu) -- (art);
\draw[varr] (art) -- (chk);
\draw[varr] (chk.north) |- node[vlab, pos=0.25, left, align=right]{yes $\cdot$ re-plan /\\ next step} (plan.west);
\draw[varr] (chk) -- node[vlab, right]{done} (final);
\end{tikzpicture}
\caption{The planner-mediated loop. The typed-artifact interface makes re-planning tractable: the planner compares structured fields, not free text, before committing to the next step. Shaded stages run on small trained specialists.}
\label{fig:planner}
\end{figure}
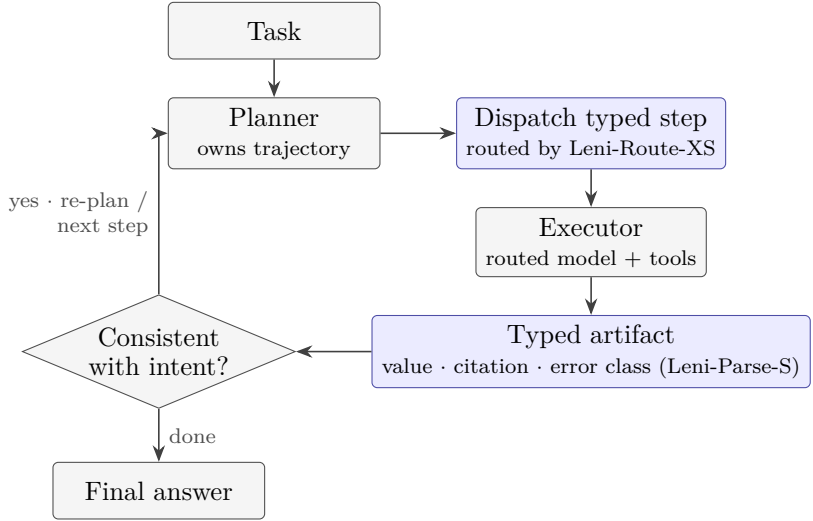

Long-horizon tool tasks fail by cascade: a misread value on step two becomes a wrong filter on step three becomes a wrong answer on step six. Here the agent splits into a planner owning the global trajectory and an executor pool specialized per step type. Executors return \emph{typed artifacts} (value, citation, confidence flag, error class) rather than raw text: Leni-Parse-S converts raw tool output into typed fields, and Leni-Route-XS selects the best model per step (fast/cheap for classification, frontier reasoning for multi-hop synthesis, strong grounding for vision) across Anthropic and OpenAI model families. The planner inspects each artifact against the original task and re-plans, inserting recovery steps or course corrections, before committing to the next step (Figure~\ref{fig:planner}). In our GAIA run the planner re-planned on ${\sim}38\%$ of tasks, mostly to recover from failed sources.

\section{Experimental Methodology}
\label{sec:method}

All evaluations use \leni{}'s production configuration (the same harness serving user tasks across spreadsheets, documents, presentations, and analytics), with no benchmark-specific prompt or harness changes. We report exact counts alongside percentages, Wilson 95\% confidence intervals for all proportions, and two-proportion $z$-tests where comparisons are made.

\subsection{SpreadsheetBench Verified}
\textbf{Data:} the 400-task expert-annotated Verified subset of SpreadsheetBench \cite{ma2024spreadsheetbench}. Tasks span formula computation, filtering, lookups, text processing, and multi-sheet operations. \textbf{Protocol:} pass@1, single attempt, no retries or cherry-picking; every cell in the designated answer region must exactly match the golden file (values read with \texttt{data\_only=True}, compared after type coercion). \textbf{Conditions:} executor Claude Opus 4.6 (Anthropic API); read-back verification by Leni-Cell-S; sandboxed code execution; adaptive extended thinking; unmodified production spreadsheet system prompt. Evaluated 16--20 April 2026. \textbf{Baseline:} the 80.25\% bare-model figure is the leaderboard's Claude Opus 4.6 entry (March 2026, minimal three-line prompt), not a rerun under our harness; the comparison is unpaired and treated accordingly.

\subsection{BullshitBench v2}
\textbf{Data:} 100 professionally framed questions containing fabricated premises across five domains (Software Engineering 40; Finance, Legal, Medical, Physics 15 each), using 13 deception techniques. \textbf{Protocol:} the benchmark's three-judge panel (GPT-4o, Gemini, and a Claude-class evaluator) scores each response 0--2; the panel score is the mean of the three judges, and a response counts as a correct rejection when the mean is at least 1.5. A 2 requires identifying the incoherence, making it central, and declining to answer within the false frame. \textbf{Conditions:} two configurations differing only in the frontier base model (Sonnet 4.6; Opus 4.6), triage on Leni-Triage-S in both. Evaluated 24 March 2026. \textbf{Caveats stated up front:} $n{=}100$; judge-based scoring with one intra-family judge; leaderboard entries (132 configurations at evaluation time) are raw API calls, so cross-entry rankings compare a system to models and are reported for context only, not as like-for-like.

\subsection{GAIA validation}
\textbf{Data:} GAIA validation split \cite{mialon2023gaia}, 165 questions (Level~1: 53; Level~2: 86; Level~3: 26). \textbf{Protocol:} exact match on final answer, no partial credit. \textbf{Conditions:} planner on Claude Opus 4.6; executors routed across Claude Opus 4.6 / Sonnet 4.6 / Haiku 4.5 and OpenAI models by Leni-Route-XS; typed-artifact extraction by Leni-Parse-S; per-step adaptive extended thinking. \textbf{Caveats stated up front:} validation answers are public, so a web-searching agent could in principle retrieve them. We ran a scripted retrieval audit over all 803 stored trajectories (results and sensitivity bound in \S\ref{sec:gaia-results}; script in the artifact bundle). The hidden test set has not yet been submitted to. Validation results here support the \emph{decomposition}, not a leaderboard claim.

\subsection{Contamination}
\label{sec:contamination}
Three contamination surfaces need separate treatment. First, base-model pretraining: we cannot rule it out for any publicly hosted benchmark; it affects the base and system arms symmetrically and is common to all published results. Second, GAIA validation answers are public and retrievable by a web-searching agent; we address this with a scripted trajectory audit and a conservative sensitivity bound (\S\ref{sec:gaia-results}). Third, specialist post-training. The training data is Leni's production corpus, three years of enterprise back-office tasks and user accept/correct/reject judgments collected independently of any benchmark campaign, with synthetic augmentation derived from production scenario patterns. The authors attest that no benchmark items, and no corpora constructed to imitate the benchmarks, were used in any training set; production work encounters situations of the same broad kind (spreadsheet edits, premise-checking, tool chains) but not the benchmarks' tasks. We ran the audit's question-side scan (word $8$/$13$-gram containment of every evaluation item, flag at ${\geq}3$ shared 8-grams or any 13-gram) over the production user-message corpus that sources the user-eval training signal: 30,104 messages against 365 evaluation items. Four messages flagged, matching three items (two GAIA validation questions, one DRACO problem) at 50--112 shared 8-grams, i.e.\ near-verbatim. All four trace to sessions from 2--17 March 2026 in which benchmark questions were manually entered during internal testing of the product, predating the recorded campaigns; three of the four originate from one internal account, and zero matches occur in organic customer traffic. Internal test accounts are excluded from training extraction (author attestation; the four flagged rows are published under pseudonymous internal-account labels, with underlying identifiers available to reviewers for verification), leaving zero evaluation items in the training-signal corpus. The scan script, thresholds, per-pair flags, and report are in the artifact repository; the workbook-trace and tool-trajectory components of the corpus remain to be swept with the same script.

\paragraph{Evaluation practice.} Reported results come only from full frozen runs. The released export preserves the complete run record, including development and superseded-configuration runs alongside the reported ones; configurations were not modified in response to results within a campaign, and no benchmark item influenced a prompt or a training set. The full run record, including unfavorable runs (a superseded L2 run at 62/86, two degraded L1 runs, and the pre-final SpreadsheetBench configurations), is preserved in the released export, which is the strongest evidence we can offer that results were not selected after the fact. The one reporting error we found, the mixed-selection 77.6\% GAIA figure, arose in aggregation, not in running, and is corrected in \S\ref{sec:gaia-results}.

\section{Results}
\label{sec:results}

\subsection{Headline: total system uplift, with uncertainty}

\begin{table}[h]
\centering\small
\resizebox{\textwidth}{!}{%
\begin{tabular}{@{}lllrrr@{}}
\toprule
\textbf{Benchmark} & \textbf{Failure mode} & \textbf{Oracle} & \textbf{Base} & \textbf{\leni{}} & \textbf{Uplift} \\
\midrule
SpreadsheetBench ($n{=}400$) & silent computation & deterministic & 80.25\% {\scriptsize[76.1, 83.9]} & 91.25\% {\scriptsize[88.1, 93.6]} & $+11.0\pp^{***}$ \\
BullshitBench / Sonnet ($n{=}100$) & confabulation & self-reflective & 91\% {\scriptsize[83.8, 95.2]} & 98\% {\scriptsize[93.0, 99.4]} & $+7\pp^{*}$ \\
BullshitBench / Opus ($n{=}100$) & confabulation & self-reflective & 87\% {\scriptsize[79.0, 92.2]} & 97\% {\scriptsize[91.5, 99.0]} & $+10\pp^{**}$ \\
GAIA validation ($n{=}165$) & cascade error & planner-mediated & ${\sim}60\%$ & 75.2\% {\scriptsize[68.0, 81.1]} & ${\sim}{+}15.2\pp$ \\
\bottomrule
\end{tabular}}
\caption{Total system uplift over the bare base model. Brackets: Wilson 95\% CIs. Significance from two-proportion $z$-tests ($^{*}p<0.05$, $^{**}p<0.01$, $^{***}p<0.001$); the GAIA base figure is an internal estimate and is not significance-tested. These comparisons measure the whole architecture (tools, multiple passes, multiple models) against raw single-pass inference; \S\ref{sec:decomposition} decomposes the total.}
\label{tab:headline}
\end{table}

Table~\ref{tab:headline} summarizes. Two comparisons are statistically unambiguous: SpreadsheetBench ($z{=}4.45$, $p<0.001$, unpaired) and BullshitBench on Opus. BullshitBench on Sonnet is nominally significant but rests on 7 discordant questions under judge-based scoring; we treat it as suggestive. The GAIA base-model figure (${\sim}60\%$) comes from an internal single run of a minimal tool harness and should be read as an estimate.

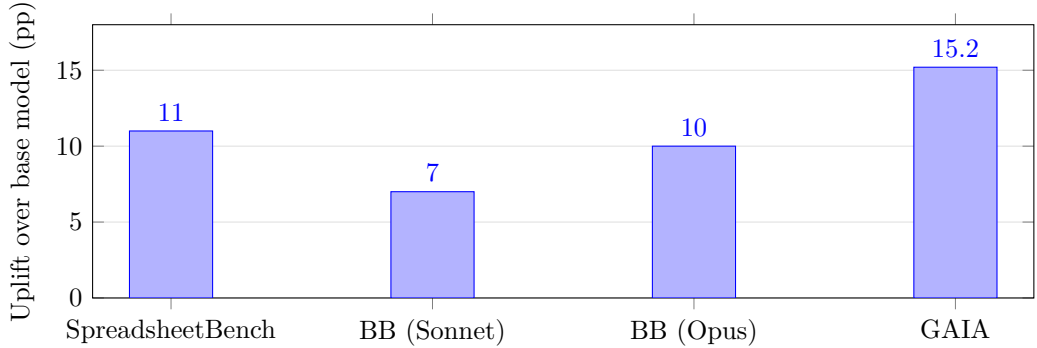
\begin{figure}[t]
\centering
\begin{tikzpicture}
\begin{axis}[
  ybar, bar width=11mm, width=0.85\textwidth, height=52mm,
  ymin=0, ymax=18, ylabel={Uplift over base model (pp)},
  symbolic x coords={SpreadsheetBench, BB (Sonnet), BB (Opus), GAIA},
  xtick=data, x tick label style={font=\small}, y tick label style={font=\small},
  ylabel style={font=\small}, ymajorgrids, grid style={black!12},
  nodes near coords, nodes near coords style={font=\small},
  every axis plot/.append style={fill=blue!25, draw=blue!50!black!70}]
\addplot coordinates {(SpreadsheetBench,11.0) (BB (Sonnet),7) (BB (Opus),10) (GAIA,15.2)};
\end{axis}
\end{tikzpicture}
\caption{Total architectural uplift over the bare base model, by benchmark. The GAIA bar is measured against an internal base-model estimate (${\sim}60\%$) and its pass@1 headline; see Table~\ref{tab:headline} for confidence intervals and \S\ref{sec:decomposition} for the decomposition of these totals into structure, specialists, and verification.}
\label{fig:uplift}
\end{figure}

We do not headline the claim that these uplifts ``exceed the gap between frontier model releases'': that comparison contrasts a system delta with a model delta on incommensurable configurations. The defensible statement is narrower and still useful: for a team holding the base model fixed, the architecture recovered more accuracy on these tasks than was available from any base-model swap we tested (Opus 4.6 vs.\ Sonnet 4.6 vs.\ GPT~5.4 as executors; internal runs).

\subsection{SpreadsheetBench: the deterministic loop, fully instrumented}
\label{sec:sb-results}

\leni{} scores 91.25\% (365/400), which at evaluation time (April 2026 snapshot) placed second among public entries behind a neurosymbolic system with a custom spreadsheet runtime (DealGlass Tetra, 94.25\%) and ahead of Nobie (91.00\%), Qingqiu/Kingsoft (89.25\%), Shortcut.ai (86.00\%), bare Claude Opus 4.6 (80.25\%), and GPT~5.4 strict (78.25\%). Every named peer is itself a scaffolded commercial agent system with verified leaderboard entries, so these are system-to-system comparisons; the bare-model figures are the leaderboard's own single-model entries and enter only the uplift calculation of Table~\ref{tab:headline}. Cell-level tasks pass at 92.4\% (254/275), sheet-level at 88.8\% (111/125); sheet-level operations (multi-region edits, cross-sheet references, structural changes) are harder, as expected.

The value of full instrumentation is the loop's confusion matrix (Table~\ref{tab:confusion}). The loop ran on 397 of 400 tasks (3 completed without triggering it; of those, 2 passed and 1 failed). Of the 397, the verifier confirmed 389 artifacts and flagged 8. All 8 flags were true errors (no observed false alarms); 6 were repaired and 2 were not. Of the 389 confirmations, 357 were correct and 32 were \emph{false confirmations}: real errors the verifier missed.

\begin{table}[h]
\centering\small
\begin{tabular}{@{}lrr@{}}
\toprule
& \textbf{Artifact correct} & \textbf{Artifact erroneous} \\
\midrule
Verifier confirmed & 357 & 32 \;(missed errors) \\
Verifier flagged & 0 \;(no false alarms) & 8 \;(6 repaired, 2 not) \\
\bottomrule
\end{tabular}
\caption{Task-level verifier confusion matrix on the 397 loop-triggering SpreadsheetBench tasks. Empirically: catch rate $c = 8/40 = 0.20$, fix rate $r = 6/8 = 0.75$, false-alarm rate $f = 0/357 = 0$ (bounded above by ${\sim}1\%$ at 95\% confidence).}
\label{tab:confusion}
\end{table}

\paragraph{What the loop caught, and what it missed.} The six rescued tasks share a signature: the recalculated values exposed defects that were invisible at write time, chiefly formulas that evaluate differently under LibreOffice than the executor assumed, cached-value mismatches, and format discrepancies that appear only after recalculation. Convergence under feedback looks stable rather than oscillatory: 26 tasks entered a second loop iteration after the agent modified its output on the first, and 22 of the 26 passed. A full run of the same configuration two days earlier (14 April) was degraded by API-server downtime and scored 356/400; the completed re-run is the reported result. The two runs disagreed on 11 tasks: 10 fail$\to$pass, consistent with recovery from the infrastructure failures, and 1 pass$\to$fail, consistent with sampling variance. This bounds run-to-run sensitivity, infrastructure faults included, at about $\pm 3\pp$. The 32 missed errors have not yet been individually coded by failure locus (comparator vs.\ recalculation-oracle vs.\ task interpretation); that coding is the immediate next step in the loop's telemetry, because it determines whether the remaining path to the neurosymbolic leader runs through a better comparator or a better executor.

Three observations. First, in the notation of \S\ref{sec:model}, the empirical operating point is $c \approx 0.20$, $r \approx 0.75$, $f \approx 0$: the verifier is conservative; it flags rarely, but its flags are precise and its fixes usually land. With $f \approx 0$, Eq.~\eqref{eq:pprime} guarantees the loop cannot hurt, and the measured rescue ($+6$ tasks net, $+1.5\pp$) matches $(1-p)\,c\,r$ within rounding. Second, the 32 missed errors quantify the cost of LLM-mediated comparison relative to symbolic enforcement: a symbolic comparator would have caught a large fraction of them, and this is a plausible mechanism for most of the 3-point gap to the neurosymbolic leader. Third, the ceiling on further gains from this loop is visible in its own telemetry: raising $c$ (a better-trained or symbolically assisted comparator) is worth up to $+8\pp$; raising $r$ is worth at most $+0.5\pp$. Instrumenting the loop tells you where to invest.

\subsection{GAIA: planner-mediated loop, and what the tier data can support}
\label{sec:gaia-results}

\paragraph{A correction, first.} An earlier company report gave 77.6\% (128/165) on this split. In preparing this paper we exported every stored GAIA trajectory from the production evaluation database (803 validation runs across seven campaigns, March--April 2026) and re-graded all of them with the official GAIA scorer; the re-grader agrees with all 218 runs that carried grades at export time (218/218). The re-grading shows the 77.6\% figure mixed selection rules across tiers: the Level~1 component (47/53) was best-of-three over repeated runs, and the Level~3 component drew on more than one campaign, while Level~2 (65/86) was a legitimate single attempt. We therefore report both quantities cleanly and retire the earlier figure.

Under a pre-specified pass@1 rule (the last completed attempt per task within the final-configuration campaigns), \leni{} scores \textbf{75.2\% (124/165)} [68.0, 81.1]: Level~1, 44/53 [70.8, 90.8]; Level~2, 65/86 [65.5, 83.4]; Level~3, 15/26 [38.9, 74.5]. Best-of-all-runs across campaigns reaches 83.0\% (137/165), a pass@$k$ quantity with $k$ varying by tier that we report for completeness, not comparison. Because Level~1 was run in full four times, GAIA also yields direct run-to-run variance: complete runs scored 44, 43, 43, and 43 of 53, with five tasks flipping outcome between the two closest runs, and two additional runs degraded by infrastructure failures (3--4 unanswered tasks each) that we report rather than exclude.

\paragraph{Retrieval audit.} Because validation answers are public, we scanned every stored trajectory for URLs of known GAIA-derived sources (the HF dataset pages, the GAIA paper, WebVoyager's \texttt{GAIA\_web.jsonl}, agents-course \texttt{metadata.jsonl}). Twenty-five of 803 trajectories, spanning 12 tasks, contain such a URL; in nearly all cases it appears inside a web-search result list rather than as a fetched page, but URL presence cannot by itself exclude use. Seven runs in the pass@1 selection are flagged, all seven scored correct. Treating every flagged correct run as a failure gives a conservative contamination-adjusted lower bound of 117/165 = 70.9\%. The audit script and per-run flags ship with the artifact bundle so the boundary between ``URL seen'' and ``answer used'' can be adjudicated by anyone from the raw trajectories.

For context, contemporaneous public figures reconstructed from per-level reports place Genspark near 75.4\%, Manus near 73.4\%, and OpenAI Deep Research near 67.4\% on this split; all three are fully scaffolded, tool-using agent systems, and their figures are self-reported. At 75.2\% pass@1, \leni{} is statistically indistinguishable from Genspark and Manus, and nominally but not significantly ahead of Deep Research ($z \approx 1.6$, $p \approx 0.12$, against a reconstructed figure). We make no superiority claim on this benchmark. The hidden test set ($n{=}300$, evaluation server) has not been submitted to; it remains the decisive surface for any cross-system claim, and we commit to reporting the result when the submission completes, whether or not it flatters the system.

What the GAIA data does establish is internal: the architecture's uplift over the ${\sim}60\%$ bare-tool-use estimate is roughly $+15\pp$ at pass@1, and the per-level telemetry (planner re-plans on ${\sim}38\%$ of tasks, mostly source-failure recovery) feeds the decomposition in \S\ref{sec:decomposition}.

\subsection{BullshitBench: sensitivity established, calibration not yet}
\label{sec:bb-results}

\leni{} reaches 98\% correct rejection on Sonnet 4.6 (mean judge score 1.963/2.0, one failure) and 97\% on Opus 4.6 (1.950/2.0, two failures), against 91\% for the best raw model configuration (Claude Sonnet 4.6 High, mean 1.87) and near-coin-flip performance for the best OpenAI and Google configurations (GPT~5.4: 48\%; Gemini~3 Pro Preview: 48\%). Domain-level rates are uniform (93--100\% across Finance, Legal, Medical, Physics, Software Engineering), consistent with a behavioral disposition induced by the firewall rather than per-domain knowledge.

\paragraph{Valid-premise control.} A firewall trained to reject premises could inflate rejection scores by over-rejecting, so sensitivity alone is not enough. We measure the other side of the boundary with DRACO, an internal rubric-graded corpus of 100 legitimate, professionally difficult, jargon-dense questions (finance, law, medicine, econometrics, technology, among others; real frameworks such as Goodman-Bacon decomposition and Callaway--Sant'Anna estimators) evaluated through the same firewall-active production harness on 16--17 April 2026, two runs per task, 201 scored runs. The system engaged substantively with every question: mean rubric score 71.3\% (median 73.5\%), no run below 10\% and no task below 20\% at task level, where a premise-rejection response would score near zero on the factual-accuracy rubrics. Zero over-rejections in 100 tasks bounds the firewall's false-positive rate on this distribution at roughly 3.6\% (95\% upper limit, rule of three). The remaining gap is distributional: DRACO questions are hard and jargon-heavy but were not constructed adversarially to mimic the benchmark's 13 deception techniques with valid premises; that matched-style control is the outstanding experiment, and until it runs, calibration against \emph{deliberately fabrication-like} valid premises remains open.

Two further qualifications. First, the $+7\pp$ Sonnet-configuration uplift is 7 questions under judge scoring with one intra-family judge, while the Opus-configuration uplift ($+10\pp$) rests on firmer ground. And ``above all 132 public configurations'' compares an agentic system against raw API calls; we report it as context for the size of the architectural effect, not as a like-for-like ranking.

\paragraph{Where the two configurations diverge.} Bucket assignments differed on three of the 100 items, and the divergences are informative about what the firewall does not control: on phys\_st\_01 (Physics, metrology) Opus validated a fabricated instrument class while Sonnet rejected it; on fin\_nn\_02 (Finance, CECL) Opus hedged without committing while Sonnet cleanly separated valid from invalid methods; on leg\_tce\_01 (Legal, contracts) Sonnet returned an empty response while Opus produced a substantive rejection. The firewall raises both configurations to near-ceiling but does not erase base-model-specific vulnerabilities; per-task judge scores for all 100 items are retained and available.

The mechanistically interesting observation survives all three qualifications: extended-thinking configurations of strong reasoners perform \emph{worse} than their standard counterparts on this benchmark, consistent with inference-time compute being spent constructing a path to an answer inside a false frame rather than questioning the frame. The firewall redirects compute toward an explicit checkpoint. Reliability comes from where the model is forced to look, not only from how much it thinks.

\section{Decomposing the Uplift}
\label{sec:decomposition}

Table~\ref{tab:decomp} decomposes total uplift by architectural layer. SpreadsheetBench is fully decomposed under the frozen harness; GAIA layer figures are from internal ablation runs (complete controlled ablation forthcoming); BullshitBench is not layer-decomposed because the firewall \emph{is} the scaffold.

\begin{table}[h]
\centering\small
\resizebox{\textwidth}{!}{%
\begin{tabular}{@{}lrrrr@{}}
\toprule
\textbf{Benchmark} & \textbf{Base} & \textbf{+ Structure} & \textbf{+ Verification loop} & \textbf{Loop-isolated} \\
\midrule
SpreadsheetBench & 80.25\% & 89.75\% \;{\scriptsize(prompt + scaffold)} & 91.25\% & $+1.5\pp$ (6 rescues) \\
GAIA & ${\sim}60\%$ & ${\sim}70\%$ {\scriptsize(+ planner--executor)} $\to$ ${\sim}74\%$ {\scriptsize(+ routing)} & 75.2\% & ${\sim}{+}1\pp$ {\scriptsize(vs.\ estimated tier)} \\
BullshitBench (Opus) & 87\% & --- & 97\% & $+10\pp$ (whole firewall) \\
\bottomrule
\end{tabular}}
\caption{Uplift by layer. Most of the total uplift comes from structure (prompting, planning, routing); the verification step's isolated contribution is small. The GAIA structure tiers are internal estimates whose selection rules were not recorded, so the GAIA loop-isolated figure is indicative only.}
\label{tab:decomp}
\end{table}

\paragraph{What the decomposition does and does not show.} The verification loop is not the majority contributor; structure is. But the loop's contribution is positionally concentrated: on SpreadsheetBench, $+1.5\pp$ is the difference between a mid-pack and a near-top result, because every system at the top has already exhausted the gains from good scaffolding. We previously believed the GAIA data showed the loop's contribution growing with chain length, as Eq.~\eqref{eq:pprime} predicts; after correcting the GAIA score to pass@1, the loop's GAIA increment (${\sim}{+}1\pp$ against an estimated structure tier) is no longer distinguishable from its spreadsheet increment. The chain-length prediction stands untested, and the controlled GAIA ablation needed to test it is the natural companion experiment to the swap ablation below.

\paragraph{Who observes matters: the specialist-swap ablation.} Holding the loop structure fixed and swapping the observe/compare stage from the trained specialist back onto the frontier model that generated the artifact: on SpreadsheetBench the rescue count falls from 6 tasks to 2 (the generator, asked to verify cells it just wrote, tends to rationalize them rather than flag them), and on BullshitBench correct rejection falls by 4--5$\pp$ (the specialist emits a hard accept/reject per claim where the general model continues a fluent answer). Two boundaries on this evidence: the swaps are single runs from the internal engineering evaluation of the specialist mix, covering Cell-S and Triage-S only (the Parse-S and Route-XS swaps have not been run), and the design lacks a third verifier condition, an independent frontier model from a different provider that did not generate the artifact, which is the cell that would separate \emph{independence} from \emph{specialization}. Within those limits, the results fit documented self-assessment biases \cite{panickssery2024selfrecognition,huang2024cannot} and support what we consider the paper's most consequential claim: \emph{the loop's value comes from the independence and specialization of the observer}.

\subsection{Internal configuration tiers, and the baselines still missing}
\label{sec:baselines}

Comparing the full system to a bare base model measures product uplift, not mechanism. Scaffold-to-scaffold comparison is not itself missing: the public leaderboard peers on SpreadsheetBench and the GAIA agent systems of \S\ref{sec:gaia-results} are all scaffolded, tool-using products, so the competitive results already compare architectures against architectures. What those comparisons cannot do is isolate mechanism, because public systems differ in base model, tools, and compute simultaneously. Two internal tiers provide partial isolation by holding the model, tools, and sandbox fixed: on SpreadsheetBench, the prompt-plus-self-validation configuration (89.75\%) is a same-model scaffold with no external oracle, and on GAIA the planner--executor tier (${\sim}70\%$) and routing tier (${\sim}74\%$) separate structure from verification. The remaining gap is a controlled same-model comparison against generic scaffolds, ReAct-style \cite{yao2023react}, CRITIC-style \cite{gou2024critic}, and best-of-$n$ at matched token budget, which would establish whether the loop beats brute-force sampling at equal compute; this isolates mechanism rather than fairness and is noted in \S\ref{sec:limitations}.

\section{Cost and Latency}
\label{sec:cost}

Per-step verification is only viable if the verifier is cheap. Per call, the specialists serve at a small fraction of the cost of the same operation on a frontier model (internal serving estimates: ${\sim}0.1\times$ for Leni-Cell-S and Leni-Triage-S, ${\sim}0.05\times$ for Leni-Parse-S, ${\sim}0.02\times$ for Leni-Route-XS, all quantized to 4~bits), which is what makes it economical to run the compare stage on all 397 loop-triggering SpreadsheetBench tasks. End-to-end latency overhead of the loop has not been measured for publication and belongs in the artifact release. On GAIA, routing by Leni-Route-XS \emph{reduces} net cost: cheap steps route to small models, funding extended reasoning on hard steps; internal estimates attribute ${\sim}4\pp$ of GAIA accuracy to routing at net-negative cost. Small trained specialists are what make ``verify every step'' an affordable default rather than a luxury.

\section{Discussion}
\label{sec:discussion}

\paragraph{Reliability is mostly an architecture problem.} For a fixed base model, the architecture recovered $+7$ to ${\sim}{+}15\pp$ across three unrelated failure modes, and the decomposition attributes this to three separable mechanisms: scaffolding/structure (largest share), specialist staffing (necessary for the loop's value; cheap), and the verification checkpoint itself (small, positionally decisive). Teams should invest in that order, but not skip the third: at the top of a leaderboard, or at the tail of a reliability SLA, the checkpoint is where the remaining failures are.

\paragraph{Reliability as enterprise infrastructure.} The components that look expensive as per-application engineering amortize differently when viewed as a layer. One recalculation oracle serves every spreadsheet task the system will ever run; one typed-artifact interface and one router serve every tool chain; the specialists serve at $0.02$--$0.1\times$ frontier cost and occupy the same loop stages everywhere. The right cost comparison is to test infrastructure and continuous integration in software engineering: a fixed investment justified by the tail of the distribution, which is where regulated back-office work concentrates its risk. A consumer chat product would struggle to justify a deterministic oracle or a post-trained verifier; a system whose outputs enter financial close, diligence, or contract review cannot justify their absence. This is why we treat the pattern as core infrastructure for high-stakes enterprise AI rather than a feature of any single application.

\paragraph{The oracle sets the ceiling; the observer sets the floor.} A commodity deterministic oracle (LibreOffice recalculation) leaves only a 3-point gap to a full neurosymbolic system at a fraction of the engineering cost, and our confusion matrix localizes that gap in the LLM-mediated comparison stage (32 missed errors). Where re-execution is possible, prefer it, and consider symbolically assisted comparison. Where it is not, an independent specialized observer still outperforms self-assessment by the generator; the swap ablation and the prior literature agree on this point.

\paragraph{Why more thinking is not enough.} Extended-thinking reasoners can do \emph{worse} on premise validation: compute is spent building a path to an answer inside a false frame. A checkpoint that forces the model to look at the frame beats additional compute spent inside it. This is the practical content of Huang et al.'s negative result \cite{huang2024cannot} for enterprise deployment: self-correction needs an external or structurally independent signal, and the architecture's job is to supply one.

\section{Limitations and Threats to Validity}
\label{sec:limitations}

\begin{itemize}[leftmargin=1.5em]
  \item \textbf{Vendor evaluation.} This is an evaluation of \leni{} by \leni{}'s builders. We mitigate with public benchmarks, exact-match protocols where available, frozen configurations, full counts, and stated CIs, but independent replication is the real remedy; we will provide evaluation transcripts to qualified researchers on request.
  \item \textbf{Limited repeated-run coverage.} The headline SpreadsheetBench and BullshitBench results are single scored runs. Measured run-to-run sensitivity where repeats exist: GAIA Level~1 complete runs scored 44/43/43/43 of 53 (5 tasks flipped between adjacent runs), and a SpreadsheetBench run interrupted by API downtime and its completed re-run disagreed on 11 of 400 tasks (10 attributable to the infrastructure failures). Differences smaller than ${\sim}3\pp$ should be treated as unresolved.
  \item \textbf{GAIA validation, not test; corrected headline.} Validation answers are public and the hidden test set has not been submitted to. The scripted retrieval audit (\S\ref{sec:gaia-results}) brackets the pass@1 result at [70.9\%, 75.2\%] depending on how flagged trajectories are adjudicated. The previously reported 77.6\% mixed best-of-$k$ and pass@1 selection and is superseded. Cross-system GAIA statements in this paper are context, not claims.
  \item \textbf{Training exclusion audited on the question corpus; two components remain.} The published $n$-gram sweep of the production message corpus found 4/30,104 matches, all adjudicated as internal manual benchmark testing with identifiers published for verification (\S\ref{sec:contamination}). Remaining on attestation: that internal test accounts are excluded from training extraction, and the sweeps of the workbook-trace and tool-trajectory corpus components.
  \item \textbf{No controlled same-model scaffold ablation.} The paper's competitive comparisons are already scaffold-to-scaffold (public agent systems on both SpreadsheetBench and GAIA), but ReAct-style, CRITIC-style, and best-of-$n$ configurations with the \emph{same} base model at matched tool access and token budget have not been run (\S\ref{sec:baselines}); how much of the loop's advantage over brute-force sampling survives compute matching is therefore unmeasured.
  \item \textbf{Competitor figures are reconstructions.} GAIA competitor overall scores are weighted estimates from per-level reports at a point in time; rankings shift. All rank statements in this paper are dated snapshots (April 2026), not standing claims.
  \item \textbf{Valid-premise control is observational, not adversarially matched.} The DRACO control (\S\ref{sec:bb-results}) bounds over-rejection at ${\lesssim}3.6\%$ on legitimate expert-level questions, but its items were not styled after the benchmark's deception techniques; a matched-style control with deliberately fabrication-like valid premises remains to be run.
  \item \textbf{Judge-based scoring.} The BullshitBench panel spans three providers but includes one judge from the same model family as the systems under test; intra-family preference is documented \cite{panickssery2024selfrecognition,zheng2023judging}.
  \item \textbf{Preliminary ablations.} The specialist-swap results cover two of four specialists, come from internal single runs, and lack the independent-generalist verifier condition; the GAIA layer decomposition is likewise from internal single runs. They motivate, but do not yet establish, the ``independent observer'' hypothesis.
  \item \textbf{Proprietary components.} Specialist weights, the production training corpus, and the production prompts are not released. The base models, sizes, and full post-training recipe are documented (\S\ref{sec:mix}), which permits method-level replication on public data, but not verification of our checkpoints; we flag this as the main obstacle to independent verification of the mechanism as opposed to the scores.
\end{itemize}

\section{Conclusion}

Across three benchmarks stressing three unrelated failure modes, a production agent architecture (verification loops staffed by lightweight post-trained specialists, on top of strong scaffolding) produces large total uplift over its frontier base model, with the two largest comparisons statistically unambiguous. Decomposition, not the headline numbers, is the durable contribution: structure supplies most of the gain; the verification checkpoint supplies a small, positionally decisive remainder; and preliminary evidence suggests the checkpoint only pays when the observer is independent of the generator. The verifier confusion matrix we report turns loop design from folklore into measurement: catch rate, fix rate, and false-alarm rate are estimable quantities that tell a team exactly where the next point of reliability will come from.

For teams building enterprise agents where a confidently wrong answer has real cost, the actionable guidance is concrete and ranked: invest first in structure (planning, routing, typed interfaces), then staff observation with a small model that did not generate the artifact, then give the loop the strongest oracle the task admits: re-execution where possible, structured re-evaluation where not. And instrument the loop: its own telemetry will tell you whether to believe it. Built this way, the verification layer stops being an application feature and becomes the reliability substrate of enterprise AI, a fixed piece of infrastructure, amortized across every task the system serves, that makes high-stakes automation defensible in the first place.

\section*{Artifact Availability and Run Configuration}

All results in this paper were computed from a CSV export of the production evaluation database: per-run records for SpreadsheetBench (1,299 runs over 400 tasks across four configurations, with per-task pass/fail and instruction type), BullshitBench (500 runs over 100 items, of which 200 carry per-judge scores, panel means, and bucket assignments), GAIA (803 validation runs over 165 tasks across seven campaigns, with stored final answers per trajectory), and the DRACO valid-premise control (309 runs over 100 tasks with per-criterion rubric evaluations). The GAIA re-grading script implements the official scorer and reproduces all 218 database-resident grades exactly; it ships with the export. The complete bundle (run records, audit scripts, and reports, including unfavorable and superseded runs) is publicly released at \url{https://github.com/arnabdastidar/leni-agent-evals}; the loop-action telemetry of Table~\ref{tab:confusion} currently exists in the run transcripts rather than as structured fields and is being extracted for the release. Specialist weights and the production training corpus are proprietary, but the method is documented to replication depth: Qwen3 bases (0.6B/1.7B/4B, Apache-2.0), distillation SFT from a Claude Opus 4.6 teacher, RLVR/GRPO against deterministic checkers (Cell-S, Parse-S, Route-XS) or SFT-then-DPO on preference pairs (Triage-S), QLoRA iteration with full-fine-tune release checkpoints, 4-bit serving, and shadow deployment before in-loop promotion.

Run configuration: SpreadsheetBench used Claude Opus 4.6 via the Anthropic API with sandboxed code execution and adaptive extended thinking, the unmodified production xlsx prompt, evaluated 16--20 April 2026. BullshitBench used the production harness on Claude Sonnet 4.6 and Claude Opus 4.6, evaluated 24 March 2026 under the benchmark's three-judge panel and bucket rule. GAIA used a Claude Opus 4.6 planner with executors routed across Claude Opus 4.6, Sonnet 4.6, Haiku 4.5, and OpenAI models, with per-step adaptive extended thinking. Sampling parameters and harness commit identifiers are not reported here and will accompany the run-log release.

\section*{Conflict of Interest and Authorship}

The corresponding author is the chief executive and founder of Leni Inc., which develops the system evaluated here; the Leni team built the production system, ran the evaluation campaigns, and trained the specialist models. All claims, analyses, and errors are the responsibility of the corresponding author. AI writing assistance was used in revising and tightening parts of this manuscript. This work has not been independently replicated; the evaluation counts, per-task outcomes, and audit scripts are publicly released at \url{https://github.com/arnabdastidar/leni-agent-evals}.

\end{document}